\documentclass[english,aip,jcp,amsmath,amssymb,reprint]{revtex4-2}
\usepackage[T1]{fontenc}
\usepackage[utf8]{inputenc}   
\usepackage{color}
\usepackage{babel}
\usepackage{textcomp}
\usepackage{amssymb}
\usepackage{graphicx}
\usepackage{hyperref}
\usepackage{dcolumn}
\usepackage{mciteplus}
\usepackage{amsmath}
\usepackage{physics}

\preprint{AIP/123-QED}
\begin{document}

\title{Photoluminescence Line Shapes of Nanocrystals: Contributions from First- and Second-Order Vibronic Couplings}

\author{Kaiyue Peng}
\email{kaiyue_peng@berkeley.edu}
\affiliation{Department of Chemistry, University of California, Berkeley, California, 94720, United States}
\affiliation{Materials Sciences Division, Lawrence Berkeley National Laboratory, Berkeley, California, 94720, United States}

\author{Bokang Hou}
\email{bkhou@berkeley.edu}
\affiliation{Department of Chemistry, University of California, Berkeley, California, 94720, United States}

\author{Kailai Lin}
\affiliation{Department of Chemistry, University of California, Berkeley, California, 94720, United States}

\author{Caroline Chen}
\affiliation{Department of Chemistry, University of California, Berkeley, California, 94720, United States}


\author{Hendrik Utzat}
\affiliation{Department of Chemistry, University of California, Berkeley, California, 94720, United States}
\affiliation{Materials Sciences Division, Lawrence Berkeley National Laboratory, Berkeley, California, 94720, United States}

\author{Eran Rabani}
\email{eran.rabani@berkeley.edu}
\affiliation{Department of Chemistry, University of California, Berkeley, California, 94720, United States}
\affiliation{Materials Sciences Division, Lawrence Berkeley National Laboratory, Berkeley, California, 94720, United States}
\affiliation{Fritz Haber Center for Molecular Dynamics, The Institute of Chemistry and the Institute of Applied Physics, The Hebrew University of Jerusalem, Jerusalem 91904, Israel}

\date{\today}
\begin{abstract}
We present a microscopic, parameter-free approach for computing the photoluminescence spectra of a single semiconductor nanocrystal. The method derives exciton-phonon coupling directly from the semi-empirical pseudopotential framework and systematically incorporates both diagonal and off-diagonal interactions, expanded to second-order in the phonon modes. The dipole-dipole correlation function was calculated using a Dyson expansion within the Kubo-Toyozawa formalism, enabling a consistent description of the role of pure dephasing and population-transfer on the photoluminescence spectral features. Applied to CdSe/CdS core–shell nanocrystals, the approach quantitatively reproduces experimental photoluminescence spectra over a wide temperature range, revealing that quadratic phonon couplings account for nearly half of the homogeneous linewidth above $\approx 100-150$K, while off-diagonal couplings leading to exciton thermalization play only a minor role and only as $T\rightarrow 300$K.
\end{abstract}

\maketitle

\section{Introduction}
The interplay between electronic excitations and nuclear motion governs a broad class of non-radiative dynamical processes that strongly influence both the efficiency and coherence of light emission in colloidal semiconductor nanocrystals (NCs).\cite{nozik_spectroscopy_2001, wheeler_exciton_2013, collins_non-radiative_2015, garcia_de_arquer_semiconductor_2021} Elucidating the microscopic nature of the exciton-phonon coupling (EXPC) is therefore essential for optimizing NCs for modern quantum and photonic technologies,\cite{burdov_exciton-photon_2021, peng2023polaritonic, peng2024polariton} from light-emitting diodes~\cite{shirasaki_emergence_2013} and single-photon sources~\cite{senellart_high-performance_2017} to nanoscale lasers~\cite{ledentsov_quantum_2011} and photovoltaics.\cite{kramer_colloidal_2011} Among optical observables, the single-nanocrystal photoluminescence (PL) spectrum provides a particularly sensitive and direct window into coupled electronic–nuclear dynamics and the associated transient relaxation and dephasing processes. Leveraging this sensitivity, in recent years homogeneous PL measurements supported by theory have been used to identify the dominant non-radiative relaxation pathways that govern spectral line shapes and linewidths.\cite{empedocles_photoluminescence_1996,valerini_temperature_2005, fernee_origin_2007, cho_temperature_2023, berkinsky_narrow_2023, lin2023theory,hou2025unraveling}

At cryogenic temperatures, homogeneous PL measurements of high-quality CdSe, InP, and perovskite nanocrystals exhibit highly structured spectra, consisting of a narrow zero-phonon line (ZPL), often with sub-meV full width at half maximum (FWHM) and in some cases limited only by instrumental resolution or spectral diffusion, together with one or more phonon sidebands that reflect characteristic coupling to optical vibrational modes.\cite{besombes_acoustic_2001, htoon_structure_2004, chilla_direct_2008, fu_unraveling_2018, utzat_coherent_2019, lv_exciton-acoustic_2021, cho_temperature_2023, berkinsky_narrow_2023, chandrasekaran_exciton_2023} The discrepancy between narrow lines with distinct phonon satellites at low temperatures and featureless broad lines at room temperature suggest that pure dephasing plays a central role in shaping the line shape. As the temperature increases, the ZPL broadens and merges with the acoustic and optical phonon sidebands, into a slightly asymmetric peak, with line widths that depend more markedly on temperature.\cite{bayer_temperature_2002, favero_temperature_2007, cho_temperature_2023, chandrasekaran_exciton_2023} The temperature-dependent crossover in the PL lineshape behavior indicates that, at elevated temperatures, additional relaxation channels, beyond pure dephasing, contribute to the observed spectral profile.\cite{lin2023theory}

While substantial advances have been made, achieving a unified theoretical framework for nanocrystal PL spectra across temperatures and dephasing mechanisms continues to offer important opportunities for further development. One key difficulty lies in the accurate and efficient calculation of electronic excited states and their vibronic couplings for large-scale systems, particularly in the intermediate to weak quantum confinement regimes, which play a central role in governing excited-state dynamics.\cite{jasrasaria2022simulations, li2025exciton, ghosh2025atomistically} The approach often taken is based on perturbatively expanding the exciton-phonon coupling in the lattice modes,\cite{krummheuer_theory_2002, muljarov_dephasing_2004} retaining only the lowest order (linear in the lattice modes), diagonal (in the excitonic basis) term. Such an approach has provided valuable insights into homogeneous PL line shapes,\cite{utzat_coherent_2019, lin2023theory} however, it ignores off-diagonal terms that mediate population transfer among excitonic states and quadratic couplings that account for mode mixing (such as Duschinsky rotations) and frequency shifts of each excited-state potential energy surface (PES). Consequently, at elevated temperatures, theoretical predictions often underestimate the line widths compared to experiment, and empirical broadening is necessary to account for missed channels.\cite{krummheuer_theory_2002, muljarov_dephasing_2004, lin2023theory}  

Here, we address this by explicitly including linear diagonal and off-diagonal terms and quadratic exciton-phonon coupling terms in the vibronic Hamiltonian. The magnitude of the excitons-phonon couplings is derived directly from the semi-empirical pseudopotential (SEPP) approach. We use the Kubo-Toyozawa method~\cite{kubo1955application} combined with a Dyson expansion approach for the off-diagonal couplings to calculate the dipole-dipole correlation function and analyze the contributions of pure dephasing and population-transfer on equal footing. Applied to CdSe/CdS core-shell NCs as a model system, our framework quantitatively reproduces experimental PL spectra across a wide temperature range, accurately capturing trends of line width broadening, phonon sideband positions, and ZPL-to-sideband intensity ratios. Importantly, all spectral features emerge directly from \textit{calculated} EXPC and exciton dephasing and relaxation dynamics, with no empirical parameters. Our analysis of the temperature-dependent FWHM and dephasing functions further reveals that low-temperature PL is dominated by diagonal linear couplings, while quadratic phonon couplings contribute substantially and account for nearly half of the homogeneous linewidth and dephasing rates above $\sim 100-150$K. Off-diagonal phonon couplings contribute weakly via population transfer processes, and are only noticeable at elevated temperatures. Together, our model quantitatively reproduces the observed PL line shape and its temperature dependence without empirical corrections. It establishes a direct link between the microscopic nature of EXPC and the macroscopic optical observables via exciton dynamics, offering a transferable route for modeling dephasing and coherence in quantum emitters.

\begin{figure}[h!tb]
\includegraphics[width=8cm]{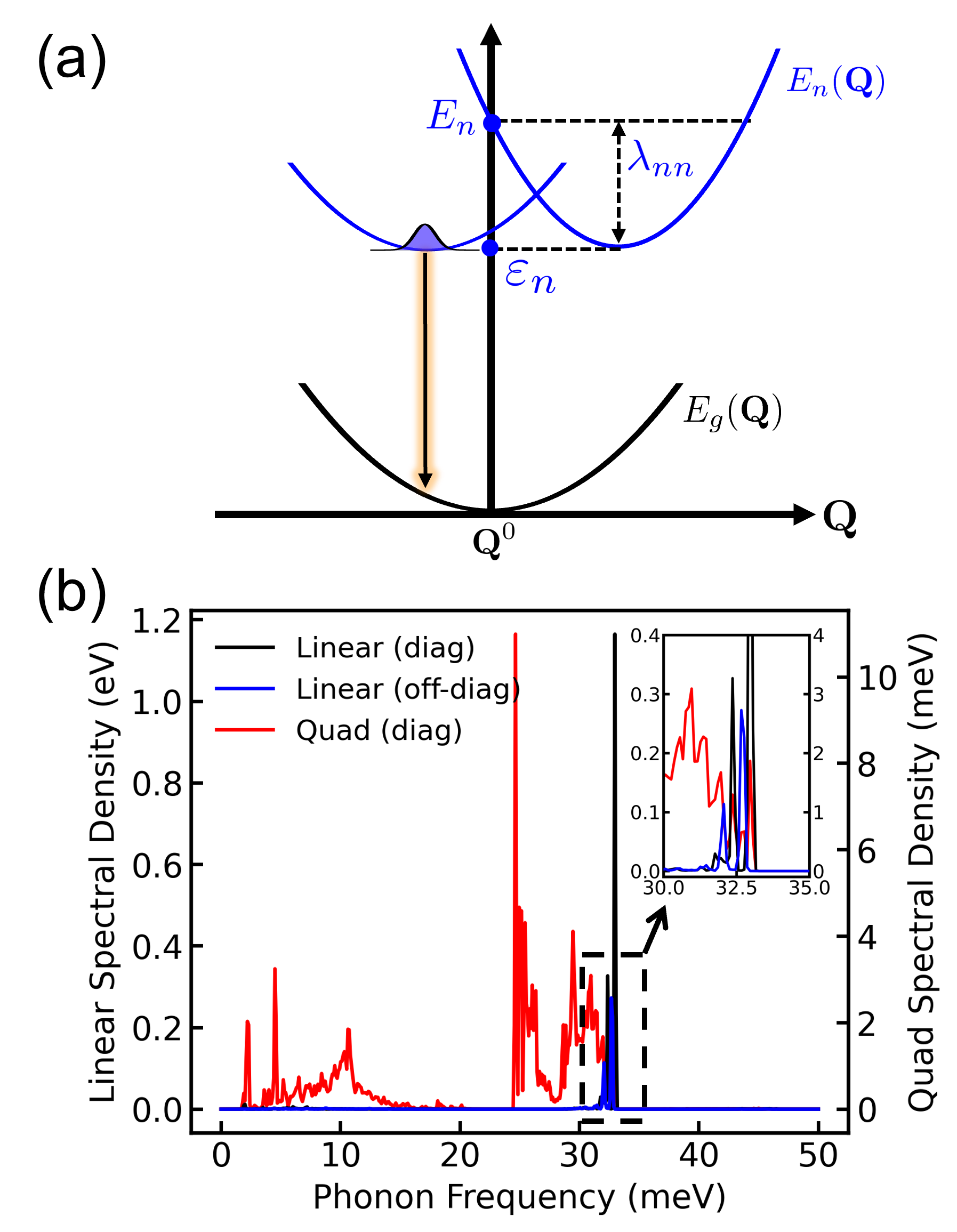}
\caption{(a) Schematic of the potential energy surfaces (PES) for the ground state (black line) and excited states (blue lines) of an NC system. The origin of the photoluminescence signal is indicated by the orange-glowing black arrow. (b) Spectral densities for linear diagonal (black solid line), linear off-diagonal (blue solid line), and quadratic diagonal (red solid line) phonon couplings of a $3$~nm diameter CdSe/3 ML CdS core-shell nanocrystal. Note that the linear couplings are plotted against the left y-axis, while the quadratic couplings are shown on the right y-axis. An inset zooms into the phonon energy range of $30-35$~meV to highlight the role of the linear off-diagonal spectral features. The spectral densities are broadened by Gaussian functions with widths of $0.05$~meV. \label{fig:fig1}}
\end{figure}

\section{Methods}
\subsection{Model Hamiltonian}
We utilize the following Hamiltonian to describe a QD weakly coupled to a monochromatic electromagnetic field within the dipole approximation:\cite{lin2023theory}
\begin{align}
\hat{H} & =\hat{H}_{\rm QD}+\mathcal{E}\left[\sum_{n}\mu_{gn}\cos\left(\omega t\right)\left|\psi_{g}\right\rangle \left\langle \psi_{n}\right|+h.c.\right].
\end{align}
In the above, $\hat{H}_{\rm QD}$ described the vibronic features of the QD within the crude adiabatic representation:\cite{jasrasaria2022simulations}
\begin{equation}
\begin{aligned}
\hat{H}_{\rm QD}&=E_{g}\left|\psi_{g}\right\rangle \left\langle \psi_{g}\right|+\sum_{n}E_{n}\left|\psi_{n}\right\rangle \left\langle \psi_{n}\right|\\&+\sum_{\alpha}\hbar\omega_{\alpha}\hat{a}_{\alpha}^{\dagger}\hat{a}_{\alpha}+\sum_{\alpha nm}V_{nm}^{\alpha}\left|\psi_{n}\right\rangle \left\langle \psi_{m}\right|\hat{q}_{\alpha}\\&+\sum_{\alpha n}W_{nn}^{\alpha}\left|\psi_{n}\right\rangle \left\langle \psi_{n}\right|\hat{q}_{\alpha}^2,
\end{aligned}
\end{equation}
where $\left|\psi_{g}\right\rangle$ is the ground state with energy $E_{g}=0$ set arbitrarily to zero, and $\left|\psi_{n}\right\rangle $ is an excited state with energy $E_{n}$. These quantities are computed at the equilibrium geometry of the NC using the semi-empirical pseudopotential methods~\cite{PhysRevB.51.17398,wang1996pseudopotential} combined with the Bethe-Salpeter equation (BSE).\cite{PhysRevB.62.4927} 

The nuclear degrees of freedom are represented by a sum of harmonic modes,  with normal mode frequencies $\omega_{\alpha}$ and mass-weighted coordinates $\hat{q}_{\alpha}=\sqrt{\frac{\hbar}{2\omega_{\alpha}}}\left(\hat{a}_{\alpha}^{\dagger}+\hat{a}_{\alpha}\right)$, obtained by diagonalizing the dynamical matrix of the NC at the equilibrium geometry using the Stillinger-Weber force field.\cite{zhou2013stillinger,plimpton1995fast} Here, $\hat{a}_{\alpha}^{\dagger}$ and $\hat{a}_{\alpha}$ are the phonon creation and annihilation operators for mode $\alpha$, respectively. $V_{nm}^{\alpha}$ and $W_{nm}^{\alpha}$ denote the first- and second-order exciton-phonon coupling strengths between the excitonic states $\left|\psi_{n}\right\rangle$ and $\left|\psi_{m}\right\rangle$ mediated by phonon mode $\alpha$. The linear couplings include both diagonal ($n=m$) and off-diagonal ($n\neq m$) terms. The quadratic couplings are restricted to diagonal terms ($n=m$), and off-diagonal quadratic couplings corresponding to Duschinsky rotations are neglected. These terms are also computed using the semi-empirical pseudopotential methods combined with the BSE.\cite{jasrasaria2023circumventing} 


Figure~\ref{fig:fig1}(a) shows the ground- and excited-state crude potential energy surfaces (PES), $E_g(\mathbf{Q})$ and $E_n(\mathbf{Q})$, as functions of the a chematic phonon mode $\mathbf{Q}=(q_{1},q_{2},\ldots)$, with equilibrium coordinate $\mathbf{Q}^{0}$. Diagonal exciton-phonon couplings displace the excited-state PES along $\mathbf{Q}$. Consequently, the PL emission associated with state $n$ lies near the energy $\varepsilon_{n}=E_{n}-\lambda_{nn}$, where the reorganization energy is $\lambda_{nn}=\sum_{\alpha}\frac{\left(V_{nn}^{\alpha}\right)^{2}}{2\omega_{\alpha}^{2}}$. 

The strength of linear and quadratic
vibronic couplings can be quantitatively analyzed using the spectral
density. The linear spectral density is given by $J_{nm}\left(\omega\right)=\sum_{\alpha}\frac{\left(V_{nm}^{\alpha}\right)^{2}}{2\omega_{\alpha}}\delta\left(\omega-\omega_{\alpha}\right)$, while quadratic spectral density is given by $\Gamma_{nm}\left(\omega\right)=\sum_{\alpha}\frac{\hbar \left(W_{nm}^{\alpha}\right)^{2}}{\omega_{\alpha}^{2}}\delta\left(\omega-\omega_{\alpha}\right)$. The linear vibronic spectral density is sharply peaked at $30$--$35$~meV, frequencies that correspond to the optical phonons of the CdSe core.~\cite{lin2023theory} The magnitudes of the off-diagonal linear couplings are about an order of magnitude smaller than the diagonal ones, indicating that the PESs are weakly coupled to each other. The quadratic spectral density is highly structured across the entire phonon density of states of the NC, spanning both acoustic and optical modes. However, it is nearly two orders of magnitude weaker than the linear spectral density. As further discussed below, the relatively weak quadratic phonon couplings nevertheless contribute significantly to the PL line width, particularly at temperatures above $150$ K.

\subsection{Photoemission}
Within linear-response theory, the emission spectrum is proportional to the Fourier transform of the dipole-dipole autocorrelation function~\cite{1130000797363520000}:
\begin{equation}
I\left(\omega\right)\propto\int_{-\infty}^{+\infty}dt\exp\left(-i\omega t\right)\left\langle \hat{\mu}\left(t\right)\hat{\mu}\left(0\right)\right\rangle, 
\end{equation}
where the transition dipole operator in the Heisenberg picture is given by
$\hat{\mu}\left(t\right)=e^{i\hat{H}_{\rm QD}t/\hbar}\hat{\mu}e^{-i\hat{H}_{\rm QD}t/\hbar}$ with
\begin{equation}
\hat{\mu}=\sum_{n}\mu_{gn} \left|\psi_{g}\right\rangle \left\langle \psi_{n}\right|+\mu_{ng}\left|\psi_{n}\right\rangle \left\langle \psi_{g}\right|, 
\end{equation}
where $\mu_{gn}$ is the transition dipole moment between ground state $g$ and excited state $n$, obtained from the semi-empirical pseudopotential methods combined with the BSE. Accordingly, the transition-dipole autocorrelation function can be expressed as:
\begin{equation}
\left\langle \hat{\mu}\left(t\right)\hat{\mu}\left(0\right)\right\rangle =\text{Tr}\left[\hat{\rho}\left(0\right)e^{i\hat{H}_{\rm QD}t/\hbar}\hat{\mu}e^{-i\hat{H}_{\rm QD}t/\hbar}\hat{\mu}\right],
\end{equation}
where $\text{Tr}\left[\cdots\right]$ denotes the trace taken over both excitonic and nuclear degrees of freedom. The density matrix at the initial time is given by $\hat{\rho}\left(0\right)=e^{-\beta\hat{H}_e}/Z_{\rm QD}$, with the partition function $Z_{\rm QD}=\mathrm{Tr}[e^{-\beta\hat{H}_{e}}]$, where $\hat{H}_{e}$ is the excited state part QD Hamiltonian:
\begin{equation}
\begin{aligned}
\hat{H}_{e}&=\sum_{n}E_{n}\left|\psi_{n}\right\rangle \left\langle \psi_{n}\right|+\sum_{\alpha}\hbar\omega_{\alpha}\hat{a}_{\alpha}^{\dagger}\hat{a}_{\alpha}\\&+\sum_{\alpha nm}V_{nm}^{\alpha}\left|\psi_{n}\right\rangle \left\langle \psi_{m}\right|\hat{q}_{\alpha}+\sum_{\alpha n}W_{nn}^{\alpha}\left|\psi_{n}\right\rangle \left\langle \psi_{n}\right|\hat{q}_{\alpha}^2.
\end{aligned}
\end{equation}
Furthermore, the transition-dipole autocorrelation function can be simplified as:
\begin{equation}
\begin{aligned}
&\left\langle \hat{\mu}\left(t\right)\hat{\mu}\left(0\right)\right\rangle \\&=\sum_{m,n\neq g}\frac{1}{Z_{\rm QD}}\mu_{gm}\mu_{gn}\left\langle \psi_{n}\right|\\&\times\text{Tr}_{\text{nu}}\left[e^{i\hat{H}_{\rm QD}(t+i\beta\hbar)/\hbar}e^{-i\hat{H}_{g}t/\hbar}\right]\left|\psi_{m}\right\rangle,
\end{aligned}
\label{eq:tru}
\end{equation}
where $\text{Tr}_{\text{nu}}\left[\cdots\right]$ denotes the trace over nuclear degrees of freedom, the ground state Hamiltonian is $\hat{H}_{g}=E_{g}\ket{\psi_g}\bra{\psi_g}+\sum_{\alpha}\hbar\omega_{\alpha}\hat{a}_{\alpha}^{\dagger}\hat{a}_{\alpha}$. 

\begin{figure*}[t]
\includegraphics[width=17cm]{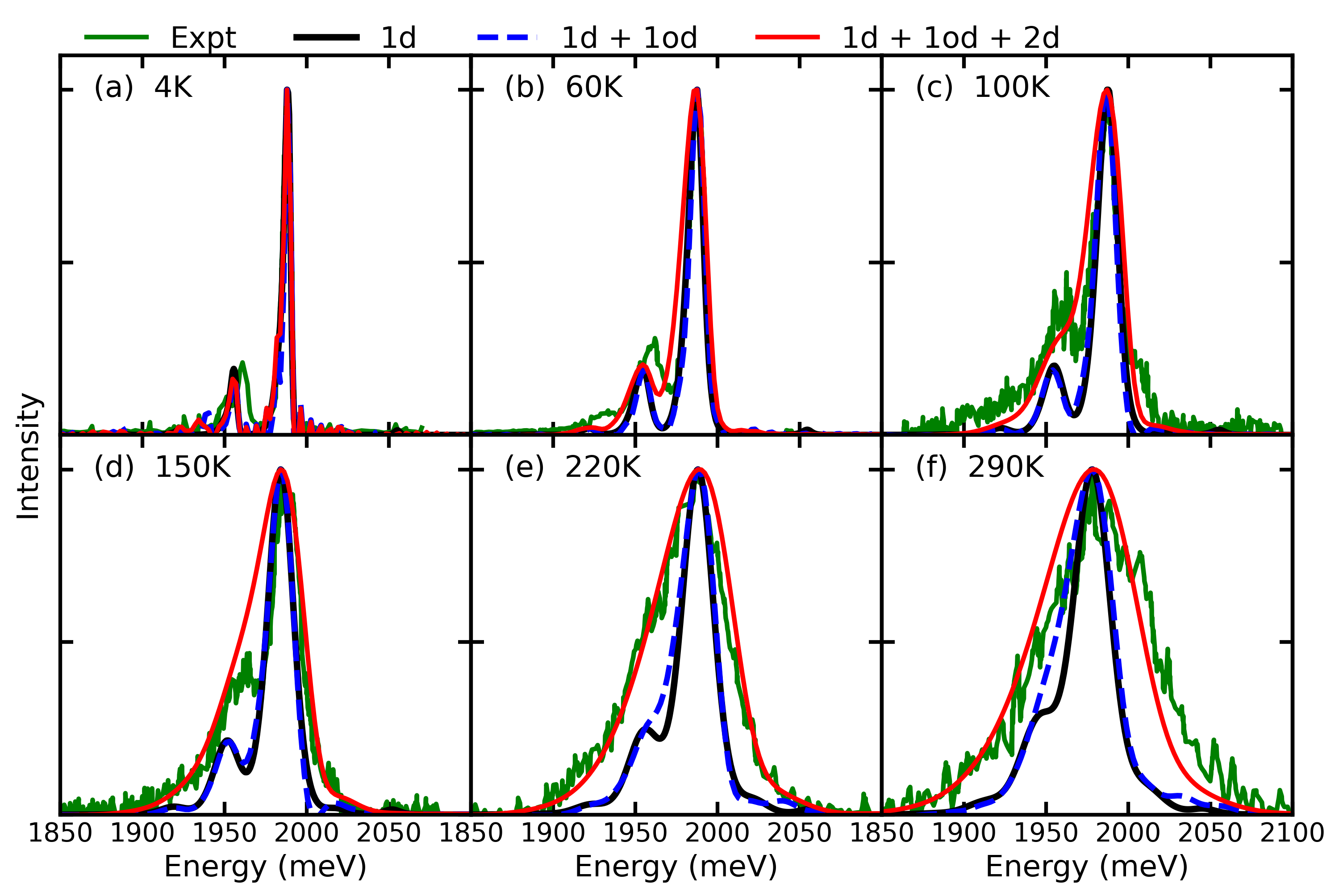}
\caption{Experimental~\cite{lin2023theory} and simulated photoluminescence spectra of a $3$~nm diameter CdSe/$3$~ML CdS core-shell nanocrystal measured at temperatures from $4$K to $290$K. Experimental spectra are shown as green solid lines. Simulated spectra obtained with (i) linear diagonal phonon couplings only (black solid lines, labeled as "1d"), (ii) both linear diagonal and linear off-diagonal phonon couplings (blue dashed lines, labeled as "1d + 1od"), and (iii) both linear and quadratic phonon couplings (red solid lines, labeled as "1d + 1od + 2d") are presented for comparison.\label{fig:fig2}}
\end{figure*}

To compute the PL line shape as given by Eq.~\eqref{eq:tru}, we partition the QD Hamiltonian $\hat{H}_{\rm QD}$ into a zeroth order part:
\begin{equation}
\begin{aligned}
\hat{H}^{0}&=E_{g}\left|\psi_{g}\right\rangle \left\langle \psi_{g}\right|+\sum_{n}E_{n}\left|\psi_{n}\right\rangle \left\langle \psi_{n}\right|\\&+\sum_{\alpha}\hbar\omega_{\alpha}\hat{a}_{\alpha}^{\dagger}\hat{a}_{\alpha}+\sum_{\alpha n}V_{nn}^{\alpha}\left|\psi_{n}\right\rangle \left\langle \psi_{n}\right|\hat{q}_{\alpha}\\&+\sum_{\alpha n}W_{nn}^{\alpha}\left|\psi_{n}\right\rangle \left\langle \psi_{n}\right|\hat{q}_{\alpha}^2, 
\end{aligned}
\end{equation}
and the interaction,
\begin{equation}
\hat{V}_{\text{off}}=\sum_{\alpha n\neq m}V_{nm}^{\alpha}\left|\psi_{n}\right\rangle \left\langle \psi_{m}\right|\hat{q}_{\alpha}.   
\end{equation}
Using these definitions, we rewrite the QD time-evolution operator (in the complex plane) using the interaction picture:\cite{}
\begin{equation}
\begin{aligned}
 e^{i\hat{H}_{\rm QD} z/\hbar}&=e^{i\left(\hat{H}^{0}+\hat{V}_{\text{off}}\right)z/\hbar}\\
 &=\widetilde{\mathcal{T}}_{\gamma} \exp{\left[\frac{i}{\hbar} \int_{\gamma(0\to z)} d \tau \hat{V}_{\text{off}}(\tau)\right]} e^{i \hat{H}^{0} z/\hbar},
\label{eq:cumu_expan}
\end{aligned}
\end{equation}
where $z = t + i \beta \hbar$, $\hat{V}_{\text{off}}\left(\tau\right)=e^{i\hat{H}^{0}\tau/\hbar}\hat{V}_{\text{off}}e^{-i\hat{H}^{0}\tau/\hbar}$, and the corresponding integral is carried out over an L-shape contour, following the path $\gamma: 0\rightarrow t \rightarrow t+i\beta\hbar$. $\widetilde{\mathcal{T}}_{\gamma}$ is the time order operator for the contour $\gamma$.

\begin{figure*}[t]
\includegraphics[width=17cm]{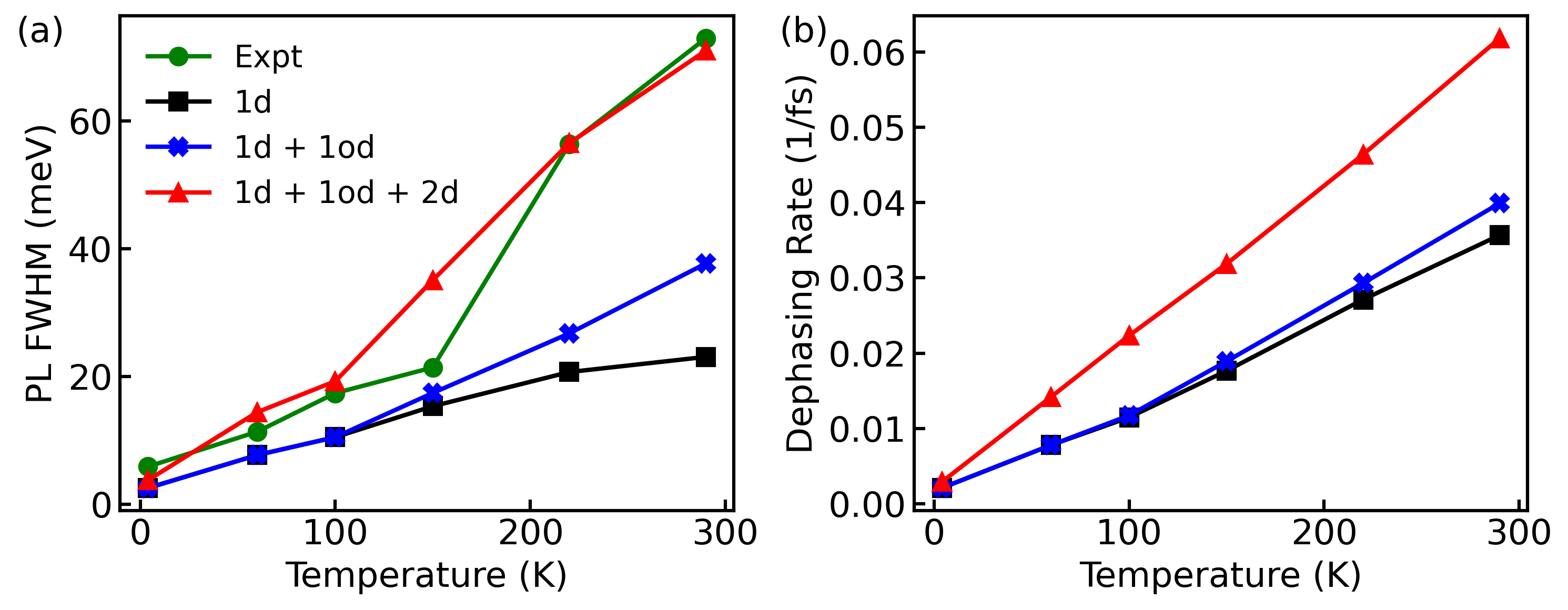}
\caption{(a) Full width at half maximum (FWHM) of the homogeneous PL spectrum as a function of temperature for the results shown in Figure~\ref{fig:fig2}. (b) Simulated dephasing rates as a function of temperature under various exciton-phonon coupling terms.\label{fig:fig3}}
\end{figure*}

Next, we expand the propagator appearing in Eq.~\eqref{eq:cumu_expan} to second order in the off-diagonal exciton-phonon coupling: 
\begin{equation}
\begin{aligned}
&\exp{\left[\frac{i}{\hbar} \int_{\gamma(0\to z)} d \tau \hat{V}_{\text{off}}(\tau)\right]}\\&\approx1+\left(\frac{i}{\hbar}\right)\int_{\gamma(0\to z)}dz_{1}\hat{V}_{\text{off}}\left(z_{1}\right)\\&
- \frac{1}{\hbar^2}\int_{\gamma(0\to z)}dz_1 \int_{\gamma(0\to z_1)}dz_2 \hat{V}_{\text{off}}(z_2)\hat{V}_{\text{off}}(z_1)\\
&\equiv 1 + K_1 + K_2.
\end{aligned}
\end{equation}
Substituting this expression into the transition-dipole autocorrelation function (Eq.~\eqref{eq:tru}) gives:
\begin{equation}\label{eq:dd_simp1}
\begin{aligned}
&\left\langle \hat{\mu}\left(t\right)\hat{\mu}\left(0\right)\right\rangle \\&\approx\sum_{m,n\neq g}\frac{1}{Z_{\rm QD}}\mu_{gm}\mu_{gn}\left\langle \psi_{n}\right|\text{Tr}_{\text{nu}}\left\{\right.\left.\left[1+K_1+K_2\right]\right.\\&\left.e^{i\hat{H}^{0}(t+i\beta\hbar)/\hbar}e^{-i\hat{H}_{g}t/\hbar}\right\} \left|\psi_{m}\right\rangle.
\end{aligned}
\end{equation}
The traces over the nuclear degrees of freedom over each term can now be evaluated exactly using the Kubo-Toyozawa method.\cite{kubo1955application,egorov_vibronic_1998} To reduce the complexity we further include only diagonal matrix elements in evaluating the dipole-dipole correlation function in Eq.~\eqref{eq:dd_simp1}, leading to  a more standard form of the emission lineshape:\cite{ma2015forster,moix2015forster}
\begin{equation}
I\left(\omega\right)\propto\sum_{n\neq g}\left|\mu_{gn}\right|^{2}\int_{-\infty}^{+\infty}dte^{-i\omega t} F_{n}\left(t\right)
\label{eq:spectr_depha},
\end{equation}
where the dephasing function is
\begin{equation}
\begin{aligned}
F_{n}(t) =\left\langle \psi_{n}\right|\text{Tr}_{\text{nu}}\left\{\left[1+K_2\right]e^{i\hat{H}^{0}(t+i\beta\hbar)/\hbar}e^{-i\hat{H}_{g}t/\hbar}\right\} \left|\psi_{n}\right\rangle,
\label{eq:dephasing}
\end{aligned}
\end{equation}

It is worthy of mentioning two limits of Eq.~\eqref{eq:dephasing}. When only diagonal phonon couplings are include, namely for $\hat{V}_{\text{off}}=0$, Eq.~\eqref{eq:dephasing} is exact. In addition, for purely linear diagonal phonon couplings, the conventional second-order cumulant expansion method provides an exact description of the dephasing function and is simpler compared with the Kubo-Toyozawa method~\cite{lin2023theory, skinner1986pure, ma2015forster} (see Supporting Information for more details).

\begin{figure*}[t]
\includegraphics[width=17cm]{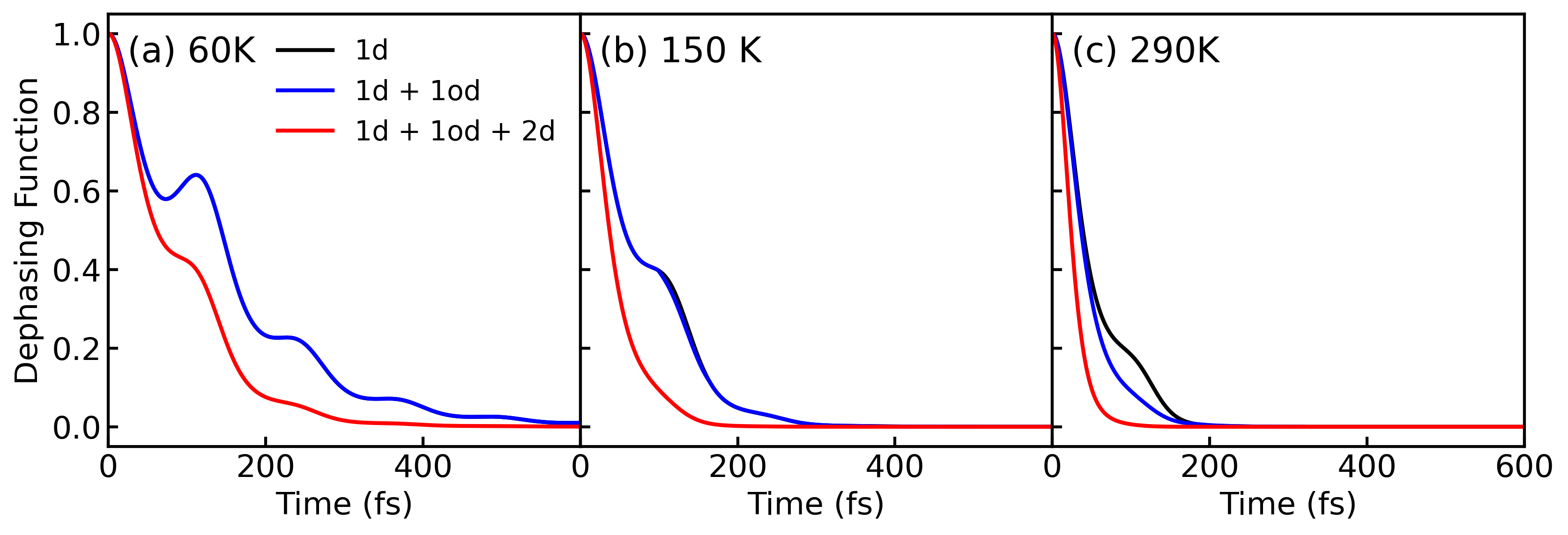}
\caption{Calculated dephasing function $\left\langle F_{n}\left(t\right)\right\rangle$ for the lowest bright excitonic state of a $3$~nm diameter CdSe/$3$~ML CdS core-shell nanocrystal, computed for the different exciton-phonon coupling orders and for three temperatures. At temperatures below $150$~K, the dephasing functions obtained with only diagonal linear exciton-phonon couplings (black solid lines) closely overlaps with the dephasing function obtained using both diagonal and off-diagonal linear phonon couplings (blue solid lines).\label{fig:fig4}}
\end{figure*}

\section{Results and Discussion}
In Figure~\ref{fig:fig2}, we plot the normalized PL spectra for a 3nm CdSe core with 3 monolayers (ML) of a CdS core-shell at different
temperatures. We show both experimental results (solid green line from Ref.~\citenum{lin2023theory}) and results based on numerical solutions of Eq.~\eqref{eq:spectr_depha} with exact expression derived for the dephasing function, given by Eq.~\eqref{eq:dephasing}. The simulation emission onset is shifted to match the experiments, in order to eliminate inaccuracies in predicting the optical gap, size of the NC, and the presence of an additional ZnS shell in the measured samples which is ignored in the model results. 

Overall, the simulated spectrum (red curve) shows excellent agreement with experimental results across a wide range of temperatures. In particular, the model follows closely the spectral lineshape, the relative intensity of the different peaks, and the crossover behavior from low to high temperatures. At low temperatures, the PL line shape is primarily governed by linear diagonal phonon couplings (black curve). At temperatures around $100-150$~K, the contribution to the spectral linewidth is nearly equal comparing the contributions of the linear (black curve) and quadratic (red curve) phonon couplings terms. Furthermore, the linear off-diagonal coupling terms associated with exciton-exciton-phonon scattering (i.e., exciton thermalization, blue cureve) influence the spectrum only at elevated temperatures, and even then their effect is minor.

The asymmetry of the PL line shape originates from the phonon sideband associated with longitudinal optical (LO) phonons of CdSe and CdS around 30 meV. The theoretical results suggest that the phonon sideband merges with the main emission peak (comprising of the zero-phonon line and a few acoustic phonon sidebands~\cite{lin2023theory}) above $100$~K, whereas experimental observations indicate such merging occurs only above $150$~K. This small discrepancy could arise from the presence of an external ZnS shell which further induces pressure in the CdSe core~\cite{Gruenwald2012} and limits the atomic motion, thereby reducing the effective exciton–phonon couplings.

To further elucidate the relative contributions of different phonon coupling terms, we plot the full width at half maximum (FWHM) as a function of temperature in Figure~\ref{fig:fig3}(a), comparing experimental observations with theoretical simulations. The results reveal that the linear off-diagonal couplings leading to exciton thermalization, contribute to the spectral features only at temperatures above 150 K, consistent with the trends observed in the PL spectra. At the highest temperature studied ($300$~K), the linear off-diagonal couplings increase the spectral linewidth by $30-40$\%.  Since off-diagonal couplings mediate population transfer between excited states, their minor impact on PL line width is a direct consequence of the relatively slow (on the order of $100-1000$~fs) exciton thermalization. Consequently, for simplicity, the off-diagonal terms have often been ignored in prior calculations. In contrast, quadratic exciton-phonon couplings contribute nearly half of the total linewidth at the highest temperature examined, emphasizing the importance of including higher-order exciton–phonon interactions, even in II–VI semiconductors where such couplings are relatively weak. Our results show that phonon interactions beyond first order play a significant role in photoluminescence spectral above the crossover temperature, underscoring the necessity of incorporating higher-order exciton-=phonon terms for an accurate description of spectral features and the polaronic dynamics.

\begin{figure*}[t]
\includegraphics[width=17cm]{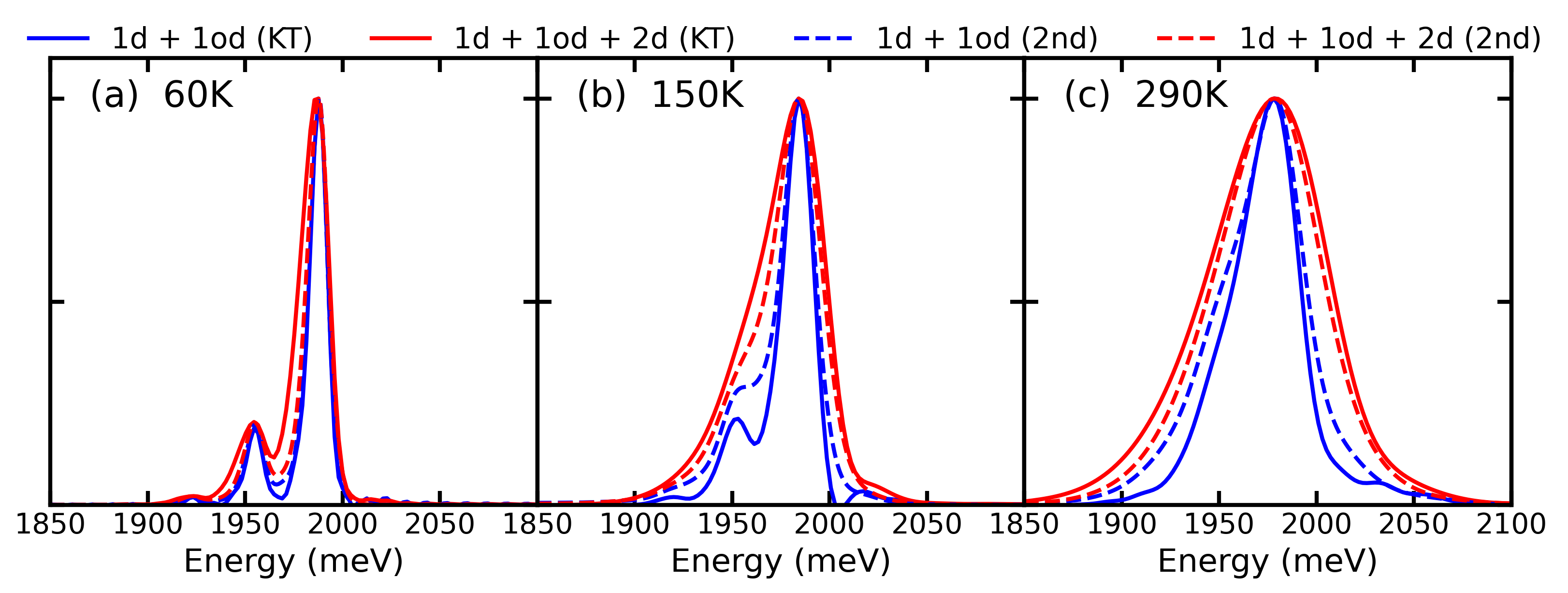}
\caption{Homogeneous PL spectra at (a) $60$K, (b) $150$K, (c) $290$K obtained using the Kubo-Toyozawa method (solid lines, labeled as "KT") and the conventional second-order cumulant expansion method (dashed lines, labeled as "2nd") for different orders of phonon couplings of a $3$~nm diameter CdSe/$3$~ML CdS core-shell nanocrystal.\label{fig:fig5}}
\end{figure*}

To further support this, we analyze the transient behavior of the dephasing function (cf., Eq.~\ref{eq:dephasing}), as shown in Figure~\ref{fig:fig4}.  We focus on the dephasing function $\left\langle F_{n}\left(t\right) \right\rangle $ of the lowest bright exciton and extract the dephasing rates by fitting the dephasing function to an exponential form, $\left\langle F_{n}\left(t\right)\right\rangle \sim\exp\left[-t/\tau\right]$. The dephasing rates are shown in  Figure~\ref{fig:fig3}(b) as a function of temperature and exhibit an approximately linear temperature dependence,  consistent with experimental observations.\cite{valerini_temperature_2005, cho_temperature_2023, berkinsky_narrow_2023, chandrasekaran_exciton_2023} Even at low temperatures, the dephasing rates arising from quadratic exciton-phonon couplings remains substantial. The pure dephasing rates induced by linear and quadratic diagonal phonon couplings constitute dominant contributions to the total dephasing rate, while the influence of population transfer mediated by off-diagonal phonon couplings is comparatively less pronounced.

As a final side note, we provide an assessment of the conventional second-order cumulant expansion in comparison to the exact Kubo-Toyozawa approach.  Figure~\ref{fig:fig5} presents the simulated PL spectra for both approaches. The results indicate that conventional second-order cumulant expansion tends to overestimate the influence of off-diagonal couplings while underestimating the contribution from quadratic diagonal couplings.  Consequently, the overall spectral linewidth and features are well described by the cumulant approximation, which can be used (with caution) instead of the more cumbersome Kubo-Toyozawa method. 


\section{Conclusion}
We have established a fully predictive theoretical framework based on a parameterized Hamiltonian to quantitatively reproduce the temperature-dependent PL spectra of CdSe/CdS core–shell NCs, without reliance on empirical fitting. Our atomistic approach, which rigorously treats both linear and quadratic exciton–phonon coupling terms, successfully captures key spectral features, including line width broadening, phonon sideband positions, and the relative intensities of the ZPL and side bands, across a broad temperature range. Through systematic analysis of the FWHM and dephasing behavior, we reveal the significance of quadratic phonon couplings, which contribute nearly half of the homogeneous line width and dephasing rate, particularly above $100$K. Moreover, we show that linear off-diagonal exciton–phonon couplings become increasingly significant at elevated temperatures (above $200$K) and is essential for accurately capturing FWHM behavior. Our unified framework provides a rigorous foundation for interpreting spectral features and dephasing mechanisms in semiconductor NCs, offering a compelling pathway for future investigations into their optical and electronic properties, particularly those associated with exciton–phonon couplings.

\section{Supporting Information}
The supplementary material comprises of the nanocrystal configuration, nanocrystal excited state calculation, exciton-phonon coupling calculation, Kubo-Toyozawa formalism, and conventional second-order cumulant expansion.

\begin{acknowledgements}
We would like to thank Professors Moungi Bawendi and David Limmer for fruitful discussions. This work was supported by the National Science Foundation Division of Chemistry, under the Chemical Theory, Models and Computational Methods (CTMC) program, grant number CHE-2449564 and the U.S. National Science Foundation Science and Technology Center (STC) for Integration of Modern Optoelectronic Materials on Demand (IMOD) under Cooperative Agreement No. DMR-2019444.  This research used resources of the National Energy Research Scientific Computing Center, a DOE Office of Science User Facility supported by the Office of Science of the U.S. Department of Energy under Contract No. DE-AC02-05CH11231 using NERSC award BES-ERCAP0032503. ER would like to thank the Israel Science Foundation for support (No. 4085/25). 
\end{acknowledgements}

\section*{Author Declarations}
\subsection*{Conflict of Interest Statement}
The authors have no conflicts to disclose.
\subsection*{Author Contributions}
Kaiyue Peng and Bokang Hou contributed equally to this work. Kaiyue Peng: Data curation (lead); Formal analysis (lead); Software (lead), Methodology (lead); Conceptualization (equal); Validation (equal); Writing – original draft (lead). Bokang Hou: Data curation (lead); Formal analysis (lead); Software (lead), Methodology (lead); Conceptualization (equal); Validation (equal); Writing – original draft (lead). Kailai Lin: Formal analysis (equal); Methodology (equal); Conceptualization (equal); Writing – original draft (lead). Caroline Chen: Formal analysis (equal); Software (equal); Methodology (equal). Hendrik Utzat: Data curation (equal); Conceptualization (equal); Writing/Review and Editing (equal). Eran Rabani: Funding Acquisition (lead); Supervision (lead); Writing/Review and Editing (lead); Methodology (equal); Project Administration (lead); Conceptualization (equal); Writing – original draft (equal).

\section*{Data Availability Statement}
The data that support the findings of this study are available
from the corresponding authors upon reasonable request.
\bibliography{main}
\end{document}


\title{Supporting Information: Photoluminescence Line Shapes of Nanocrystals: Contributions from First- and Second-Order Phonon Couplings}

\author{Kaiyue Peng}
\email{kaiyue_peng@berkeley.edu}
\affiliation{Department of Chemistry, University of California, Berkeley, California,
94720, United States}
\affiliation{Materials Sciences Division, Lawrence Berkeley National Laboratory, Berkeley, California,
94720, United States}

\author{Bokang Hou}
\email{bkhou@berkeley.edu}
\affiliation{Department of Chemistry, University of California, Berkeley, California, 
94720, United States}

\author{Kailai Lin}
\affiliation{Department of Chemistry, University of California, Berkeley, California, 
94720, United States}

\author{Caroline Chen}
\affiliation{Department of Chemistry, University of California, Berkeley, California, 
94720, United States}


\author{Hendrik Utzat}
\affiliation{Department of Chemistry, University of California, Berkeley, California, 
94720, United States}

\author{Eran Rabani}
\email{eran.rabani@berkeley.edu}
\affiliation{Department of Chemistry, University of California, Berkeley, California,
94720, United States}
\affiliation{Materials Sciences Division, Lawrence Berkeley National Laboratory, Berkeley, California,
94720, United States}
\affiliation{Fritz Haber Center for Molecular Dynamics, The Institute of Chemistry and the Institute of Applied Physics,
The Hebrew University of Jerusalem, Jerusalem 91904, Israel}
\maketitle

\section{Nanocrystal configuration}
The configurations of the core-shell CdSe/CdS NCs in theoretical calculations were constructed by first cutting the desired core size from bulk wurtzite CdSe geometry, and then adding several monolayers of CdS shell. The 3 nm CdSe core with a 3 monolayers (MLs) CdS shell QD contains a total of 1197 Cd atoms, 252 Se atoms, and 945 S atoms. 

The Stillinger--Weber force field~\cite{zhou2013stillinger} was utilized to determine the equilibrium structure and normal vibrational modes using LAMMPS.~\cite{plimpton1995fast} Following minimization, the outermost layer was replaced by ligands with a modification of the pseudopotential to represent the passivation layer.\cite{rabani1999electronic}

\section{Nanocrystal excited state calculation}
We used the same pseudopotential function as described in ref.~\citenum{jasrasaria2022simulations} for Cd, Se, and S elements.
The electron and hole single-particle states, near the top of the valence band and the bottom of the conduction band, were generated using the filter-diagonalization technique on real space grid bases.~\cite{toledo2002very} The exciton states were constructed as linear combinations of the product of electron-hole wavefunctions with coefficients obtained by solving the Bethe-Salpeter equation (BSE) within the static screening approximation.~\cite{PhysRevB.62.4927} The $n$-th exciton wavefunction can be written as:
\begin{equation}
|\psi_{n}\rangle=\sum_{ai}c_{ai}^{n}|\phi_{a}\rangle\otimes|\phi_{i}\rangle,
\end{equation}
where $\phi_{a}$ and $\phi_{i}$ are the $a$-th and $i$-th electron and hole states, respectively. $\left\{ c_{ai}^{n}\right\}$  is the BSE coefficient obtained with a static dielectric constant $\epsilon=5$. Here we use 60 hole states and 80 electron states to converge the excited state calculations. 

\section{Exciton-phonon Coupling Calculation}
It can be shown directly that
\begin{align}
    V_{nm}^\alpha &= \bra{\psi_n}\left(\frac{\partial U}{\partial q_\alpha}\right)_{\mathbf{R}_0}\ket{\psi_m}\\
    W_{nm}^{\alpha\beta} &= \bra{\psi_n}\left(\frac{\partial^2 U}{\partial q_\beta\partial q_\alpha}\right)_{\mathbf{R}_0}\ket{\psi_m}
\end{align}
where $U(\mathbf{r},\mathbf{R}_{0})=\sum_{\mu}\hat{v}_{\mu}\left(|\mathbf{r}-\mathbf{R}_{\mu,0}|\right)$ is the total potential and $\hat{v}_{\mu}\left(|\mathbf{r}-\mathbf{R}_{\mu,0}|\right)$ is the constructed semi-empirical pseudopotential for each atom $\mu$. Using finite difference along each $q_\alpha$ coordinate and displace $\delta q_\alpha$
\begin{align}
    \left(\frac{\partial U}{\partial q_\alpha}\right)_{q_\alpha=0}&\approx\frac{U(+\delta q_\alpha)-U(-\delta q_\alpha)}{2\delta q_\alpha}\\
    \left(\frac{\partial^2 U}{\partial q_\alpha\partial q_\beta}\right)_{q_\alpha, q_\beta=0}&\approx\frac{U(\delta q_\alpha,\delta q_\beta)-U(\delta q_\alpha,-\delta q_\beta)-U(-\delta q_\alpha,\delta q_\beta)+U(-\delta q_\alpha,-\delta q_\beta)}{4\delta q_\alpha\delta q_\beta}.
\end{align}
If we only kept the diagonal second order couplings, then
\begin{equation}
    \left(\frac{\partial^2 U}{\partial q_\alpha^2}\right)_{q_\alpha=0}\approx\frac{U(\delta q_\alpha)-2U(q_\alpha=0)+U(-\delta q_\alpha)}{\delta q_\alpha^2}.
\end{equation}
Note that the total pseudopotential is written in the electronic and nuclear coordinates, i.e. $U(\mathbf{r},\mathbf{R})$, so we should relate the phonon displacement with the atomic displacement when performing the finite difference. Suppose we choose the energy scale of $q_\alpha$ coordinate to be $k_BT$, where the temperature $T$ is a finite difference parameter. The maximum displacement $q_\alpha^{\rm max}\equiv\delta q_\alpha = \sqrt{k_BT}/\omega_\alpha$. For the motion along a single mode, the atomic displacement is related to the normal mode by 
\begin{equation}
    R_{\mu k}^\alpha - R_{\mu k, 0} = E_{\mu k}^\alpha\frac{q_\alpha}{\sqrt{M_\mu}},
\end{equation}
where $R_{\mu k}^\alpha$ labels the coordinate of atom $\mu$ along mode $\alpha$ and direction $k$ with $k\in\left\{x,y,z\right
\}$, $M_\mu$ is the mass of the atom $\mu$. Then we have
\begin{equation}
    R_{\mu k}^\alpha = R_{\mu k, 0} + E_{\mu k}^\alpha\frac{\delta q_\alpha}{\sqrt{M_\mu}} = R_{\mu k, 0} + E_{\mu k}^\alpha\frac{\sqrt{k_BT}}{\sqrt{M_\mu}\omega_\alpha},
\end{equation}
and the total potential energy
\begin{equation}
    U(\pm\delta q_\alpha) = U\left(\left\{R_{\mu k, 0} \pm E_{\mu k}^\alpha\frac{\sqrt{k_BT}}{\sqrt{M_\mu}\omega_\alpha}\right\}\right).
\end{equation}

\section{Kubo-Toyozawa Formalism}
As shown in the main text, the emission spectrum is given by
\begin{equation}
I\left(\omega\right)\propto\sum_{n\neq g}\left|\mu_{gn}\right|^{2}\int_{-\infty}^{+\infty}dte^{-i\omega t} F_{n}\left(t\right)
\label{eq:spectr_depha},
\end{equation}
where the dephasing function is defined as
\begin{equation}
\begin{aligned}
F_{n}(t) =\left\langle \psi_{n}\right|\text{Tr}_{\text{nu}}\left\{\left[1+K_2\right]e^{i\hat{H}^{0}(t+i\beta\hbar)/\hbar}e^{-i\hat{H}_{g}t/\hbar}\right\} \left|\psi_{n}\right\rangle,
\label{eq:dephasing}
\end{aligned}
\end{equation}
and can be evaluated exactly within the Kubo-Toyozawa formalism. To proceed, we split the dephasing function into two contributions, $F_n(t)\equiv F_n^{(0)}(t) + F_n^{(2)}(t)$, where
\begin{align}\label{eq:F^0}
    F_n^{(0)}(t) &= \left\langle \psi_{n}\right|\text{Tr}_{\text{nu}}\left\{e^{i\hat{H}^{0}(t+i\beta\hbar)/\hbar}e^{-i\hat{H}_{g}t/\hbar}\right\} \left|\psi_{n}\right\rangle\\
        F_n^{(2)}(t) &= \left\langle \psi_{n}\right|\text{Tr}_{\text{nu}}\left\{K_2e^{i\hat{H}^{0}(t+i\beta\hbar)/\hbar}e^{-i\hat{H}_{g}t/\hbar}\right\} \left|\psi_{n}\right\rangle. \label{eq:F^2}
\end{align}
Before evaluating $F_n^{(0)}(t)$ and $F_n^{(2)}(t)$, we rewrite the zeroth-order Hamiltonian as
\begin{align}\label{eq:H0_new}
    \hat{H}^0 = \hat{H}_g\ket{\psi_g}\bra{\psi_g} + (\varepsilon_n+\hat{H}_n)\ket{\psi_n}\bra{\psi_n}
\end{align}
where
\begin{align}
    \hat{H}_g &= E_g+\sum_\alpha\frac{\hat{p}^2_\alpha}{2}+\frac{1}{2}\omega_\alpha^2\hat{q}_\alpha^2\\
    \hat{H}_n &= \sum_\alpha \frac{\hat{p}_\alpha^2}{2}+\frac{1}{2}\omega_{\alpha,n}^2\left(\hat{q}_\alpha+\frac{V_{nn}^\alpha}{\omega_{\alpha,n}^2}\right)^2\\
    \varepsilon_n = &E_n-\lambda_{nn}, \hspace{0.5cm}\omega_{\alpha,n}^2=\omega_\alpha^2+2W_{nn}^\alpha.
\end{align}
The polaron energy $\varepsilon_n$ corresponds to the minimum of the 
$n$-th excited-state PES (Fig.~1) and is shifted from the vertical excitation energy $E_n$ by the reorganization energy $\lambda_{nn}=\sum_\alpha {(V_{nn}^\alpha)^2}/{(2\omega_{\alpha,n}^2)}$. The normal mode frequencies for each PES, $\omega_{\alpha,n}$ are renormalized by the second-order couplings $W_{nn}^\alpha$.

Using Eq.~\ref{eq:H0_new}, the zeroth-order dephasing function in Eq.~\eqref{eq:F^0} can be rewritten as 
\begin{align}
    \begin{split}
        F_n^{(0)}(t) &= \text{Tr}_{\text{nu}}\left\{e^{-\beta\hat{H}_n}e^{i\hat{H}_nt/\hbar}e^{-i\hat{H}_{g}t/\hbar}\right\}e^{i\varepsilon_nt/\hbar}\\
    &= \int d\mathbf{Q}_0\bra{\mathbf{Q}_0}e^{-\beta\hat{H}_n}e^{i\hat{H}_nt/\hbar}e^{-i\hat{H}_{g}t/\hbar}\ket{\mathbf{Q}_0} e^{i\varepsilon_nt/\hbar}\\
    &= \int d\mathbf{Q}_0d\mathbf{Q}_1d\mathbf{Q}_2\bra{\mathbf{Q}_0}e^{-\beta\hat{H}_n}\ket{\mathbf{Q}_1}\bra{\mathbf{Q}_1}e^{i\hat{H}_nt/\hbar}\ket{\mathbf{Q}_2}\bra{\mathbf{Q}_2}e^{-i\hat{H}_gt/\hbar}\ket{\mathbf{Q}_0}e^{i\varepsilon_nt/\hbar}.
    \end{split}
\end{align}
Because $\hat{H}_n$ and $\hat{H}_g$ are quadratic in the nuclear coordinates $\mathbf{Q}$, the resulting multidimensional integral is Gaussian and can be evaluated analytically by introducing the combined coordinates
$\mathbf{q}_\alpha=\begin{pmatrix}Q_{0,\alpha} & Q_{1,\alpha} & Q_{2,\alpha}\end{pmatrix}^{\mathsf T}$
\begin{align}\label{eq:cf_gi}
    \begin{aligned}
    &\int d\mathbf{Q}_0d\mathbf{Q}_1d\mathbf{Q}_2 \bra{\mathbf{Q}_0}e^{-\beta\hat{H}_n}\ket{\mathbf{Q}_1}\bra{\mathbf{Q}_1}e^{i\hat{H}_nt/\hbar}\ket{\mathbf{Q}_2}\bra{\mathbf{Q}_2}e^{-i\hat{H}_gt/\hbar}\ket{\mathbf{Q}_0}\\
    &=\mathcal{N}(t)\int \exp{\sum_\alpha -\frac{1}{2} \mathbf{q}_\alpha^{\top} \mathbf{A}_\alpha \mathbf{q}_\alpha+\mathbf{J}^{\top}_\alpha \mathbf{q}_\alpha+c_\alpha(t)} \mathrm{d}^{3M}\mathbf{q}\\
    &= (2 \pi)^{3M / 2}\mathcal{N}(t)(\prod_\alpha\operatorname{det} \mathbf{A}_\alpha)^{-1 / 2} \exp \left[\frac{1}{2} \mathbf{J}_\alpha^{\top} \mathbf{A}^{-1}_\alpha \mathbf{J}_\alpha+c_\alpha(t)\right].
    \end{aligned}
\end{align}
with
\begin{align*}
    c_\alpha(t) = \dfrac{i\,(V_{nn}^\alpha)^2}{\hbar\omega_{\alpha,n}^3}\Bigl[\tan\bigl(\tfrac{\omega_{\alpha,n}t}{2}\bigr)+i\,\tanh\bigl(\tfrac{\beta\hbar\omega_{\alpha,n}}{2}\bigr)\Bigr]
\end{align*}
\begin{align*}
    \mathbf{J}_\alpha^\top=\frac{V_{nn}^\alpha}{\hbar\omega_{\alpha,n}}\begin{pmatrix}
-{ \tanh\bigl(\frac{\beta \hbar\omega_{\alpha,n}}{2}\bigr)}&
{i\tan\bigl(\tfrac{\omega_{\alpha,n}t}{2}\bigr)-\,\tanh\bigl(\tfrac{\beta \hbar\omega_{\alpha,n}}{2}\bigr)}&
{i\, \tan\bigl(\tfrac{\omega_{\alpha,n}t}{2}\bigr)}
\end{pmatrix}
\end{align*}
\begin{align*}
\mathbf{A}_\alpha=
\frac{1}{\hbar}
\resizebox{\textwidth}{!}{
$\displaystyle
\begin{pmatrix}
 -i\,\omega_{\alpha}\,\cot\bigl(\omega_{\alpha} t\bigr)\;+\;\omega_{\alpha,n}\,\coth\bigl(\beta\,\hbar\,\omega_{\alpha,n}\bigr)
&
 -\dfrac{\omega_{\alpha,n}}{\sinh\bigl(\beta\,\hbar\,\omega_{\alpha,n}\bigr)}
&
 \dfrac{i\omega_{\alpha}}{\sin\bigl(\omega_{\alpha} t\bigr)}
\\[1em]
 -\dfrac{\omega_{\alpha,n}}{\sinh\bigl(\beta\,\hbar\,\omega_{\alpha,n}\bigr)}
&
 \omega_{\alpha,n}\,\Bigl[i\,\cot\bigl(\omega_{\alpha,n} t\bigr)\;+\;\coth\bigl(\beta\,\hbar\,\omega_{\alpha,n}\bigr)\Bigr]
&
 -\,\dfrac{i\omega_{\alpha,n}}{\sin\bigl(\omega_{\alpha,n} t\bigr)}
\\[1em]
 \dfrac{i\omega_{\alpha}}{\sin\bigl(\omega_{\alpha} t\bigr)}
&
 -\,\dfrac{i\omega_{\alpha,n}}{\sin\bigl(\omega_{\alpha,n} t\bigr)}
&
 i\,\Bigl[-\,\omega_{\alpha}\,\cot\bigl(\omega_{\alpha} t\bigr)\;+\;\omega_{\alpha,n}\,\cot\bigl(\omega_{\alpha,n} t\bigr)\Bigr]
\end{pmatrix}$
}
\end{align*}

The second-order dephasing function in Eq.~\eqref{eq:F^2} can be evaluated using a similar scheme. The second-order term $K_2$ can be written as
\begin{equation}
    \begin{split}
        K_2 &= - \frac{1}{\hbar^2}\int_{\gamma(0\to z)}dz_1 \int_{\gamma(0\to z_1)}dz_2 \hat{V}_{\text{off}}(z_2)\hat{V}_{\text{off}}(z_1)\\
        &=-\frac{1}{\hbar^2}\int_{0}^t dt_1 \int_{0}^{t_1}dt_2 \hat{V}_{\text{off}}(t_2)\hat{V}_{\text{off}}(t_1)\\
        &-\frac{i}{\hbar^2}\int_{0}^{\beta\hbar}ds_1 \int_0^t dt_2 \hat{V}_{\text{off}}(t_2)\hat{V}_{\text{off}}(t+is_1) + \frac{1}{\hbar^2}\int_{0}^{\beta\hbar}ds_1 \int_{0}^{s_1}ds_2 \hat{V}_{\text{off}}(t+is_2)\hat{V}_{\text{off}}(t+is_1)\\
        &\equiv K_2^{\rm (RR)} + K_2^{\rm (RI)} + K_2^{(\rm II)}.
    \end{split}
\end{equation}
The second-order contribution to the dephasing function can then be written as
\begin{align}\label{eq:Fn^2_expand}
    \begin{aligned}
        F_n^{(2)}(t) &= \left\langle \psi_{n}\right|\text{Tr}_{\text{nu}}\left\{\left[K_2^{\rm (RR)} + K_2^{\rm (RI)} + K_2^{(\rm II)}\right]e^{i\hat{H}^{0}(t+i\beta\hbar)/\hbar}e^{-i\hat{H}_{g}t/\hbar}\right\} \left|\psi_{n}\right\rangle.
    \end{aligned}
\end{align}
The three terms in Eq.~\eqref{eq:Fn^2_expand} are evaluated analogously. For example,
\begin{align}
    \begin{aligned}
        &\left\langle \psi_{n}\right|\text{Tr}_{\text{nu}}\left\{K_2^{\rm (RR)}e^{i\hat{H}^{0}(t+i\beta\hbar)/\hbar}e^{-i\hat{H}_{g}t/\hbar}\right\} \left|\psi_{n}\right\rangle\\
        &=-\frac{1}{\hbar^2}\int_0^{t} d t_1 \int_0^{t_1} d t_2 \mathrm{Tr}_{\rm nu}\left\{e^{-\beta\hat{H}_n}e^{-i\hat{H}_gt/\hbar}e^{i\hat H_nt_2/\hbar} \hat{V}^{\text{off}}_{nl}e^{-i\hat H_l(t_2-t_1)/\hbar}\hat{V}^{\text{off}}_{ln}e^{-i\hat{H}_n(t_1-t)/\hbar}\right\}\\
        &\times e^{-i\tilde{\omega}_{nl}(t_1-t_2)}e^{i\varepsilon_n t/\hbar}e^{-\beta\varepsilon_n}.
    \end{aligned}
\end{align}
The trace over the nuclear degrees of freedom can again be expressed as a Gaussian integral and evaluated analytically:
 \begin{align}
     \begin{aligned}
          & \mathrm{Tr}_{\rm nu}\left\{e^{-\beta\hat{H}_n}e^{-i\hat{H}_gt/\hbar}e^{i\hat H_nt_2/\hbar} \hat{V}^{\text{off}}_{nl}e^{-i\hat H_l(t_2-t_1)/\hbar}\hat{V}^{\text{off}}_{ln}e^{-i\hat{H}_n(t_1-t)/\hbar}\right\}\\
        &= \int d\mathbf{Q}_0d\mathbf{Q}_1d\mathbf{Q}_2d\mathbf{Q}_3d\mathbf{Q}_4\bra{\mathbf{Q}_0}e^{-\beta\hat{H}_n}\ket{\mathbf{Q}_1}\bra{\mathbf{Q}_1}e^{-i\hat H_0t/\hbar}\ket{\mathbf{Q}_2}\bra{\mathbf{Q}_2}e^{i\hat{H}_nt_2}\ket{\mathbf{Q}_3}\\
        &\times\bra{\mathbf{Q}_3}e^{-i\hat H_l(t_2-t_1)/\hbar}\ket{\mathbf{Q}_4}\bra{\mathbf{Q}_4}e^{-i\hat{H}_n(t_1-t)/\hbar}\ket{\mathbf{Q}_0}(\mathbf{V}^{\rm off}_{nl}\cdot{\mathbf{Q}}_3)(\mathbf{V}^{\rm off}_{ln}\cdot{\mathbf{Q}}_4)\\
        &= \left[(2 \pi)^{5M / 2}\prod_{\alpha=1}^M N_\alpha(\operatorname{det} \mathbf{A}_\alpha)^{-1 / 2} \right]\exp \left[\sum_{\alpha=1}^M\frac{1}{2}\mathbf{J}^{\top}_\alpha \mathbf{A}^{-1}_\alpha \mathbf{J}_{\alpha} + c_\alpha\right]\\
    &\times\left[\left(\sum_{\alpha=1}^M V_{nl}^\alpha\left[\mathbf{A}_{\alpha}^{-1}\mathbf{J}_{\alpha}\right]_3 \right)\left(\sum_{\alpha=1}^M V_{ln}^\alpha\left[\mathbf{A}_{\alpha}^{-1}\mathbf{J}_{\alpha}\right]_4\right)+\sum_{\alpha=1}^M V_{nl}^\alpha V_{nl}^\alpha\left[\mathbf{A}_{\alpha}^{-1}\right]_{34}\right].
     \end{aligned}
 \end{align}
where
\begin{figure}[ht!]
  \centering
  \resizebox{\textwidth}{!}{$c_\alpha(t,t_1,t_2) =\dfrac{1}{\hbar}\left[
\dfrac{i\,\big(V_{ll}^{\alpha}\big)^2}{\omega_{\alpha,l}^{\,3}}\,
\tan\!\left(\frac{\omega_{\alpha,l}}{2}\,(t_1-t_2)\right)
\;+\;
\dfrac{\big(V_{nn}^{\alpha}\big)^2}{\omega_{\alpha,n}^{\,3}}
\left(
-\coth\!\big(\beta \hbar \omega_{\alpha,n}\big)
+\csch\!\big(\beta \hbar \omega_{\alpha,n}\big)
+i\left[
\tan\!\left(\frac{\omega_{\alpha,n}}{2}\,(t-t_1)\right)
+\tan\!\left(\frac{\omega_{\alpha,n}}{2}\,t_2\right)
\right]
\right)
\right]$}
\end{figure}

\begin{equation*}
\resizebox{\linewidth}{!}{$\displaystyle
\begin{aligned}
\mathbf{J}_\alpha(t,t_1,t_2)=\begin{pmatrix}
\displaystyle
\frac{i\,V_{nn}^{\alpha}\!\left[\tan\!\left(\tfrac{1}{2}\,\omega_{\alpha,n}(t-t_{1})\right)
+ i\,\tanh\!\left(\tfrac{\beta \hbar \omega_{\alpha,n}}{2}\right)\right]}{\hbar\,\omega_{\alpha,n}}
&
\displaystyle
-\frac{V_{nn}^{\alpha}\,\tanh\!\left(\tfrac{\beta \hbar \omega_{\alpha,n}}{2}\right)}{\hbar\,\omega_{\alpha,n}}
&
\displaystyle
\frac{i\,V_{nn}^{\alpha}\,\tan\!\left(\tfrac{1}{2}\,\omega_{\alpha,n} t_{2}\right)}{\hbar\,\omega_{\alpha,n}}
&
\displaystyle
\frac{i\!\left[-V_{ll}^{\alpha}\,\omega_{\alpha,n}\,\tan\!\left(\tfrac{1}{2}\,\omega_{\alpha,l}(-t_{1}+t_{2})\right)
+ V_{nn}^{\alpha}\,\omega_{\alpha,l}\,\tan\!\left(\tfrac{1}{2}\,\omega_{\alpha,n} t_{2}\right)\right]}{\hbar\,\omega_{\alpha,l}\,\omega_{\alpha,n}}
&
\displaystyle
-\frac{i\!\left[V_{ll}^{\alpha}\,\omega_{\alpha,n}\,\tan\!\left(\tfrac{1}{2}\,\omega_{\alpha,l}(-t_{1}+t_{2})\right)
+ V_{nn}^{\alpha}\,\omega_{\alpha,l}\,\tan\!\left(\tfrac{1}{2}\,\omega_{\alpha,n}(-t+t_{1})\right)\right]}{\hbar\,\omega_{\alpha,l}\,\omega_{\alpha,n}}
\end{pmatrix}
\end{aligned}
$}
\end{equation*}

\begin{equation*}
\resizebox{\linewidth}{!}{$\displaystyle
\begin{aligned}
\mathbf{A}_\alpha(t,t_1,t_2)=\frac{1}{\hbar}
\begin{pmatrix}
\omega_{\alpha,n}\!\left[i\,\cot\!\bigl(\omega_{\alpha,n}(t-t_{1})\bigr)+\coth\!\bigl(\beta\hbar \omega_{\alpha,n}\bigr)\right]
&
-\omega_{\alpha,n}\,\csch\!\bigl(\beta\hbar \omega_{\alpha,n}\bigr)
&
0
&
0
&
i\,\omega_{\alpha,n}\,\csc\!\bigl(\omega_{\alpha,n}(-t+t_{1})\bigr)
\\[6pt]
-\omega_{\alpha,n}\,\csch\!\bigl(\beta\hbar \omega_{\alpha,n}\bigr)
&
-\,i\,\omega_{\alpha}\,\cot\!\bigl(\omega_{\alpha} t\bigr)+\omega_{\alpha,n}\,\coth\!\bigl(\beta\hbar \omega_{\alpha,n}\bigr)
&
i\,\omega_{\alpha}\,\csc\!\bigl(\omega_{\alpha} t\bigr)
&
0
&
0
\\[6pt]
0
&
i\,\omega_{\alpha}\,\csc\!\bigl(\omega_{\alpha} t\bigr)
&
i\!\left(-\omega_{\alpha}\,\cot\!\bigl(\omega_{\alpha} t\bigr)+\omega_{\alpha,n}\,\cot\!\bigl(\omega_{\alpha,n} t_{2}\bigr)\right)
&
-\,i\,\omega_{\alpha,n}\,\csc\!\bigl(\omega_{\alpha,n} t_{2}\bigr)
&
0
\\[6pt]
0
&
0
&
-\,i\,\omega_{\alpha,n}\,\csc\!\bigl(\omega_{\alpha,n} t_{2}\bigr)
&
i\!\left(\omega_{\alpha,l}\,\cot\!\bigl(\omega_{\alpha,l}(t_{1}-t_{2})\bigr)+\omega_{\alpha,n}\,\cot\!\bigl(\omega_{\alpha,n} t_{2}\bigr)\right)
&
i\,\omega_{\alpha,l}\,\csc\!\bigl(\omega_{\alpha,l}(-t_{1}+t_{2})\bigr)
\\[6pt]
i\,\omega_{\alpha,n}\,\csc\!\bigl(\omega_{\alpha,n}(-t+t_{1})\bigr)
&
0
&
0
&
i\,\omega_{\alpha,l}\,\csc\!\bigl(\omega_{\alpha,l}(-t_{1}+t_{2})\bigr)
&
-\,i\!\left(\omega_{\alpha,l}\,\cot\!\bigl(\omega_{\alpha,l}(-t_{1}+t_{2})\bigr)+\omega_{\alpha,n}\,\cot\!\bigl(\omega_{\alpha,n}(-t+t_{1})\bigr)\right)
\end{pmatrix}
\end{aligned}
$}
\end{equation*}
The other two terms in Eq.~\eqref{eq:Fn^2_expand} are evaluated similarly.

\section{Conventional Second-order Cumulant Expansion}
Here we introduce the conventional second-order cumulant expansion on all EXPC terms to calculate transition dipole autocorrelation function.~\cite{doi:10.1021/j100412a013,lin2023theory}
We partition our Hamiltonian into three parts:
\begin{equation}
\hat{H}_{QD}=\hat{H}_{S}^{0}+\hat{H}_{ph}+\hat{V},
\end{equation}
where the zero-order electronic Hamiltonian $\hat{H}_{S}^{0}=E_{g}\left|\psi_{g}\right\rangle \left\langle \psi_{g}\right|+\sum_{n}E_{n}\left|\psi_{n}\right\rangle \left\langle \psi_{n}\right|$, zero-order phonon bath Hamiltonian: $\hat{H}_{ph}=\sum_{\alpha}\hbar\omega_{\alpha}\hat{a}_{\alpha}^{\dagger}\hat{a}_{\alpha}$, and exciton-phonon coupling perturbation: $\hat{V}=\sum_{\alpha nm}V_{nm}^{\alpha}\left|\psi_{n}\right\rangle \left\langle \psi_{m}\right|\hat{q}_{\alpha}+\sum_{\alpha n}W_{nn}^{\alpha}\left|\psi_{n}\right\rangle \left\langle \psi_{n}\right|\hat{q}^2_{\alpha}$. Therefore the transition dipole autocorrelation function can be written as $\left(E_{g}=0\right)$:
\begin{equation}
\begin{aligned}
&\left\langle \hat{\mu}\left(t\right)\hat{\mu}\left(0\right)\right\rangle 
\\&=\left\langle e^{i\hat{H}_{QD}t/\hbar}\hat{\mu}e^{-i\hat{H}_{QD}t/\hbar}\hat{\mu}\right\rangle 
\\&=\sum_{m,n\neq g}\frac{e^{-\beta E_{n}}}{\mathrm{Tr}\left[e^{-\beta\hat{H}_{{\rm QD}}}\right]}\mu_{gm}\mu_{gn}\left\langle \psi_{n}\right|\text{Tr}_{\text{nu}}\left[e^{-\beta\left(\hat{H}_{ph}+\hat{V}\right)}e^{i\left(\hat{H}_{S}^{0}+\hat{H}_{ph}+\hat{V}\right)t/\hbar}e^{-i\hat{H}_{ph}t/\hbar}\right]\left|\psi_{m}\right\rangle 
\\&=\sum_{m,n\neq g}\frac{e^{-\beta E_{n}}}{\mathrm{Tr}\left[e^{-\beta\hat{H}_{{\rm QD}}}\right]}\mu_{gm}\mu_{gn}e^{i\omega_{m}t}\\&\times\left\langle \psi_{n}\right|\text{Tr}_{\text{nu}}\left[e^{-\beta\left(\hat{H}_{s}^{0}+\hat{H}_{ph}+\hat{V}\right)}e^{i\left(\hat{H}_{S}^{0}+\hat{H}_{ph}+\hat{V}\right)t/\hbar}e^{-i\left(\hat{H}_{ph}+\hat{H}_{S}^{0}\right)t/\hbar}e^{\beta\left(\hat{H}_{S}^{0}+\hat{H}_{ph}\right)}e^{-\beta\hat{H}_{ph}}\right]\left|\psi_{m}\right\rangle.
\end{aligned}
\end{equation}
Then the trace of nuclear degree of freedom $\text{Tr}_{\text{nu}}\left[\bullet\right]$ can be written back to the average over a harmonic bath $\hat{H}_{ph}$ with initial time density matrix given by $e^{-\beta\hat{H}_{ph}}$:
\begin{equation}
\begin{aligned}
&\left\langle \hat{\mu}\left(t\right)\hat{\mu}\left(0\right)\right\rangle 	\\&=\sum_{m,n\neq g}\frac{e^{-\beta E_{n}}}{\mathrm{Tr}\left[e^{-\beta\hat{H}_{{\rm QD}}}\right]}\mu_{gm}\mu_{gn}e^{i\omega_{m}t}\left\langle \psi_{n}\right|\left\langle e^{i\left(\hat{H}_{S}^{0}+\hat{H}_{ph}+\hat{V}\right)\left(t+i\beta\hbar\right)/\hbar}e^{-i\left(\hat{H}_{ph}+\hat{H}_{s}^{0}\right)\left(t+i\beta\hbar\right)/\hbar}\right\rangle _{\text{ph}}\left|\psi_{m}\right\rangle.
\end{aligned}
\end{equation}
From here, we define a complex time $z=t+i\beta\hbar$. The cumulant representation of time-evolution operator  will be written as an intergal integrated over a contour in complex space, choosing the integral path $\gamma: 0\rightarrow t\rightarrow t+i\beta\hbar$: 
\begin{equation}
\left\langle \hat{\mu}\left(t\right)\hat{\mu}\left(0\right)\right\rangle =\sum_{m,n\neq g}\frac{e^{-\beta E_{n}}}{\mathrm{Tr}\left[e^{-\beta\hat{H}_{{\rm QD}}}\right]}\mu_{gm}\mu_{gn}e^{i\omega_{m}t}\left\langle \psi_{n}\right|\left\langle \widetilde{\mathcal{T}_{\gamma}}\exp\left[\frac{i}{\hbar}\int_{0}^{z}d\tau\hat{V}\left(\tau\right)\right]\right\rangle _{\text{ph}}\left|\psi_{m}\right\rangle,
\end{equation}
where $\hat{V}\left(\tau\right)=e^{i\left(\hat{H}_{S}^{0}+\hat{H}_{ph}\right)\tau/\hbar}\hat{V}e^{-i\left(\hat{H}_{S}^{0}+\hat{H}_{ph}\right)\tau/\hbar}$.
Perform a cumulant expansion to second order on $\hat{V}$:
\begin{equation}
\begin{aligned}
&\left\langle \hat{\mu}\left(t\right)\hat{\mu}\left(0\right)\right\rangle 	\\&\approx\sum_{m,n\neq g}\frac{e^{-\beta E_{n}}}{\mathrm{Tr}\left[e^{-\beta\hat{H}_{{\rm QD}}}\right]}\mu_{gm}\mu_{gn}e^{i\omega_{m}t}\\&\times
\left\langle \psi_{n}\right|\exp\left\{ \frac{i}{\hbar}\int_{\gamma(0\rightarrow z)}dz_1\left\langle \hat{V}\left(z_1\right)\right\rangle _{\text{ph}}-\frac{1}{\hbar^{2}}\int_{\gamma(0\rightarrow z; z_2<z_1)} dz_{1}dz_{2}\left[\left\langle \hat{V}\left(z_{2}\right)\hat{V}\left(z_{1}\right)\right\rangle _{\text{ph}}-\left\langle \hat{V}\right\rangle _{\text{ph}}^{2}\right]\right\} \left|\psi_{m}\right\rangle,
\end{aligned}
\end{equation}
where the terms depend on more than the second order of $\hat{V}$ have been ignored. The integral in complex space is given by:
\begin{equation}
\begin{aligned}
&-\frac{1}{\hbar^{2}}\int_{\gamma(0\rightarrow z; z_2<z_1)}dz_{1}dz_{2}\left\langle \hat{V}\left(z_{2}\right)\hat{V}\left(z_{1}\right)\right\rangle _{\text{ph}}
\\&=-\frac{1}{\hbar^{2}}\int_{0}^{t}dt_{1}\int_{0}^{t_{1}}dt_{2}\left\langle \hat{V}\left(t_{2}\right)\hat{V}\left(t_{1}\right)\right\rangle _{\text{ph}}
-\frac{i}{\hbar^{2}}\int_{0}^{\hbar\beta}ds_{1}\int_{0}^{t}dt_{2}\left\langle \hat{V}\left(t_{2}\right)\hat{V}\left(t+is_{1}\right)\right\rangle _{\text{ph}}
\\&+\frac{1}{\hbar^{2}}\int_{0}^{\hbar\beta}ds_{1}\int_{0}^{s_{1}}ds_{2}\left\langle \hat{V}\left(t+is_{2}\right)\hat{V}\left(t+is_{1}\right)\right\rangle _{\text{ph}}. 
\end{aligned}
\end{equation}
Under secular approximation, we ignore the $n\neq m$ term in $\left\langle \hat{\mu}\left(t\right)\hat{\mu}\left(0\right)\right\rangle$, then the transition dipole autocorrelation function can be given analytically as:
\begin{equation}
\left\langle \hat{\mu}\left(t\right)\hat{\mu}\left(0\right)\right\rangle \approx\sum_{n\neq g}\frac{e^{-\beta E_{n}}}{\mathrm{Tr}\left[e^{-\beta\hat{H}_{{\rm QD}}}\right]}\left|\mu_{ng}\right|^{2}e^{i\omega_{n}t}\exp\left\{ R_{n}\left(t\right)+J_{n}\left(t\right)+H_{n}\left(t\right)\right\},
\end{equation}
where:
\begin{equation}
\begin{aligned}
R_{n}\left(t\right)&=\sum_{\alpha}\left(it-\hbar\beta\right)\frac{W_{nn}^{\alpha}}{2\omega_{\alpha}}\frac{e^{\beta\hbar\omega_{\alpha}}+1}{e^{\beta\hbar\omega_{\alpha}}-1}\\J_{n}\left(t\right)&=\sum_{\alpha}\sum_{k}\frac{V_{nk}^{\alpha}V_{kn}^{\alpha}}{2\hbar\omega_{\alpha}}\left[\frac{e^{\beta\hbar\omega_{\alpha}}}{e^{\beta\hbar\omega_{\alpha}}-1}\left\{ \frac{it-\hbar\beta}{\left(\omega_{nk}-\omega_{\alpha}\right)}+\frac{e^{-i\left(\omega_{nk}-\omega_{\alpha}\right)t}e^{\left(\omega_{nk}-\omega_{\alpha}\right)\hbar\beta}-1}{\left(\omega_{nk}-\omega_{\alpha}\right)^{2}}\right\} \right.\\&+\left.\frac{1}{e^{\beta\hbar\omega_{\alpha}}-1}\left\{ \frac{it-\hbar\beta}{\left(\omega_{nk}+\omega_{\alpha}\right)}+\frac{e^{-i\left(\omega_{nk}+\omega_{\alpha}\right)t}e^{\left(\omega_{nk}+\omega_{\alpha}\right)\hbar\beta}-1}{\left(\omega_{nk}+\omega_{\alpha}\right)^{2}}\right\} \right]\\H_{n}\left(t\right)&=-\sum_{\alpha}\frac{W_{nn}^{\alpha}W_{nn}^{\alpha}}{4\omega_{\alpha}^{2}}\left[\left(2n_{\alpha}^{2}+2n_{\alpha}\right)\left(t+i\beta\hbar\right)^{2}+\left(2n_{\alpha}+1\right)\left(\frac{1-e^{-2\omega_{\alpha}\hbar\beta}e^{2i\omega_{\alpha}t}}{2\omega_{\alpha}^{2}}+\frac{it-\hbar\beta}{\omega_{\alpha}}\right)\right.\\&+\left.n_{\alpha}^{2}\left(\frac{2-e^{2\omega_{\alpha}\hbar\beta}e^{-2i\omega_{\alpha}t}-e^{-2\omega_{\alpha}\hbar\beta}e^{2i\omega_{\alpha}t}}{2\omega_{\alpha}^{2}}\right)\right], 
\end{aligned}
\end{equation}
with $n_{\alpha}=\frac{1}{(e^{\beta\hbar\omega_{\alpha}}-1)}$. The dephasing function, defined as:
\begin{equation}
\left\langle F_{n}\left(t\right)\right\rangle =e^{i\omega_{n}t}\left\langle \psi_{n}\right|\left\langle \widetilde{\mathcal{T}_{\gamma}}\exp\left[\frac{i}{\hbar}\int_{0}^{z}dz\hat{V}\left(z\right)\right]\right\rangle _{\text{ph}}\left|\psi_{n}\right\rangle,
\end{equation}
under the same approximation is given by:
$\left\langle F_{n}\left(t\right)\right\rangle\approx e^{i\omega_{n}t}\exp\left\{R_{n}\left(t\right)+J_{n}\left(t\right)+H_{n}\left(t\right)\right\}$.
\bibliography{main}